\documentclass[conference]{IEEEtran}
\IEEEoverridecommandlockouts
\usepackage[english]{babel}
\usepackage[utf8]{inputenc}
\usepackage{cite}
\usepackage{amsmath,amssymb,amsfonts}
\usepackage{algorithmic}
\usepackage{url}
\usepackage{graphicx}
\usepackage{textcomp}
\usepackage{xcolor}
\usepackage[utf8]{inputenc}

\def\BibTeX{{\rm B\kern-.05em{\sc i\kern-.025em b}\kern-.08emT\kern-.1667em\lower.7ex\hbox{E}\kern-.125emX}}

\begin{document}
	
\title{Beyond Trend Following:\\ Deep Learning for Market Trend Prediction}
\author{
    \IEEEauthorblockN{Fernando Berzal, Ph.D.}
    \IEEEauthorblockA{
	\textit{Dept. Computer Science and Artificial Intelligence} \\
	\textit{University of Granada}\\
	Granada, Spain \\
	berzal@acm.org
    }
    \and
    \IEEEauthorblockN{Alberto Garc\'\i a, CFA, CAIA, FRM, CMT}
    \IEEEauthorblockA{
	\textit{Head of Global Asset Allocation} \\
	\textit{ACCI Capital Investments SGIIC}\\
	Madrid, Spain \\
	albertogarcia@accipartners.com
    }
}
\maketitle

\markboth{ACCI Capital Investments, May 2024}
{Berzal, García: Beyond Trend Following: Deep Learning for Market Trend Prediction}

\begin{abstract}
Trend following and momentum investing are common strategies employed by asset managers. Even though they can be helpful in the proper situations, they are limited in the sense that they work just by looking at past, as if we were driving with our focus on the rearview mirror. In this paper, we advocate for the use of Artificial Intelligence and Machine Learning techniques to predict future market trends. These predictions, when done properly, can improve the performance of asset managers by increasing returns and reducing drawdowns.
\end{abstract}
	

\section{Introduction}

\IEEEPARstart{T}{rend} following or trend trading is an investment strategy based on the expectation of price movements to continue in the same direction: buy an asset when its price goes up, sell it when its price goes down. For its application, obviously, you need a particular criterion to detect when prices move in a particular direction over time. Since every investor uses his own criterion, a market trend is often just a perceived tendency within a financial market.

Traditional trend following is usually done on futures. Just follow trends on a large, diversified set of futures markets, covering major asset classes. Diversification is key: with multiple assets with low or negative correlations, you can achieve higher returns at a lower risk.

Trend following on stocks can easily yield negative returns in the short side (when prices go down). When we trade only on the long side, it does not always add any real value. Compared with a passive index ETF, trend following requires additional work and creates potential risks, yet it does not always yield actual benefits.

Cole Wilcox and Eric Crittenden \cite{wilcox2005} proposed the use of an all-time high as the entry criteria: buy on all time high and sell at a trailing stop set at 10 times the 40-day ATR [average true range], using a large stock universe. Trend following on single stocks, or a few of them, however, is not attractive for the risk you have to assume.

Standard trend following is not expected to work with stocks, since their correlation is too high. But momentum investing does. When a stock price goes up for a while, the likelihood of rising higher is greater than the likelihood of falling. Likewise, a stock going up faster than other stocks is likely to keep going up faster than other stocks. This is the momentum effect, known at least since the 1960's \cite{levy1967}.

Why does momentum investing work? According to academic studies \cite{jegadeesh1993}, just because people overreact to information. One explanation is that people who buy past winners and sell past losers temporarily move prices. An alternative explanation is that the market underreacts to information on short-term prospects but overreacts to information on long-term prospects. In any case, choosing the top past performers can yield positive returns. Additional safeguards can also be employed, such as not investing in stocks during bear markets.

For instance, Andreas Clenow \cite{clenow2015} employs the following trading rules on a weekly basis: rank stocks on volatility-adjusted momentum (using an exponential 90-day regression, multiplied by its coefficient of determination), calculate position sizes (targeting a daily move of 10 basis points), check the index filter (S\&P 500 above its 200-day moving average), and build your portfolio. Individual stocks are disqualified when they are below their 100-day moving average or have experienced a gap over 15\%. When, in the weekly portfolio rebalancing, a stock is no longer in the top 20\% of the S\&P 500 ranking or fails to meet the qualification criteria (moving average and gap), it is sold. It is replaced by other stocks only if the index is in a positive trend. Twice per month, position sizes are also rebalanced to control risk.

What is the difference between trend following and momentum investing? Apart from the fact that both are valid investment strategies for different situations and asset classes, the key factor is that trend following just employs the asset's past returns, i.e., its time series momentum. In contrast, momentum investing compares an asset's momentum to the momentum of other assets. Whereas trend following is essentially autoregressive, momentum investing takes (a limited) context into account. 

A common drawback of both trend following and momentum investing is that they reap benefits after the current trend is already underway. Can we do better? Most investment professionals might say that no, you cannot reliably predict changes in market trends.

The ACCI·ON project started with the goal of using Artificial Intelligence techniques to predict trends in financial markets, including both equity markets and fixed-income markets. As we will see, including context in our Machine Learning models is essential to predict future market trends.

\section{Investment Philosophy}

Instead of trying to design a fully-automated algorithmic trading decision, the goal of the ACCI·ON project was, from its inception, the design of a tool to support the work of asset managers, not to replace them. Human asset managers are in charge of their assets under management [AuM] and they should feel fully responsible for their management decisions.

\subsection{Risk Indicators}

Deciding when to change the composition of a portfolio is one of the key decisions an asset manager has to make. Proper timing is important, yet it is really hard for a human being to determine when to buy/sell assets given the overabundance of signals and the deluge of information available at his fingertips. Hence, ACCI decided to base its investment strategy on risk indicators that help asset managers time buying/selling decisions.

The basic idea of a risk indicator in this context is that a single number summarizes the current market situation, indicating the probability of a severe drawdown in the market of interest (e.g. S\&P 500, NASDAQ 100, investment-grade bonds, or high-yield bonds). When such an indicator surpasses a predefined threshold, the asset manager can take a more risk-seeking position in his portfolio. When the indicator falls below a given value, the asset manager should cover his positions and defensively switch to a more risk-averse portfolio.

Given the prior experience of ACCI managers, the risk indicators are real-valued numbers, between -1 and +1. When the risk indicator is negative, asset managers should be defensive with respect to risks in the market the indicator is designed for. When an indicator approaches -1, the probability of a severe drawdown in its market tends to one. When the risk indicator is positive, asset managers could take a more positive attitude towards the market trend. In the limit, when the risk indicator approaches +1, the probability of a severe drawdown tends to zero. Of course, risk indicator models are probabilistic and some uncertainty is always present.

As we mentioned before, asset managers are always in charge. They can modulate their risk exposure by establishing different thresholds for changing their portfolio composition. When their outlook is optimistic, they can set a lower threshold for their positive portfolio. When their outlook is pessimistic, they can set a higher threshold for abandoning their defensive portfolio.

\begin{figure*}
\centering
\includegraphics[width=\textwidth]{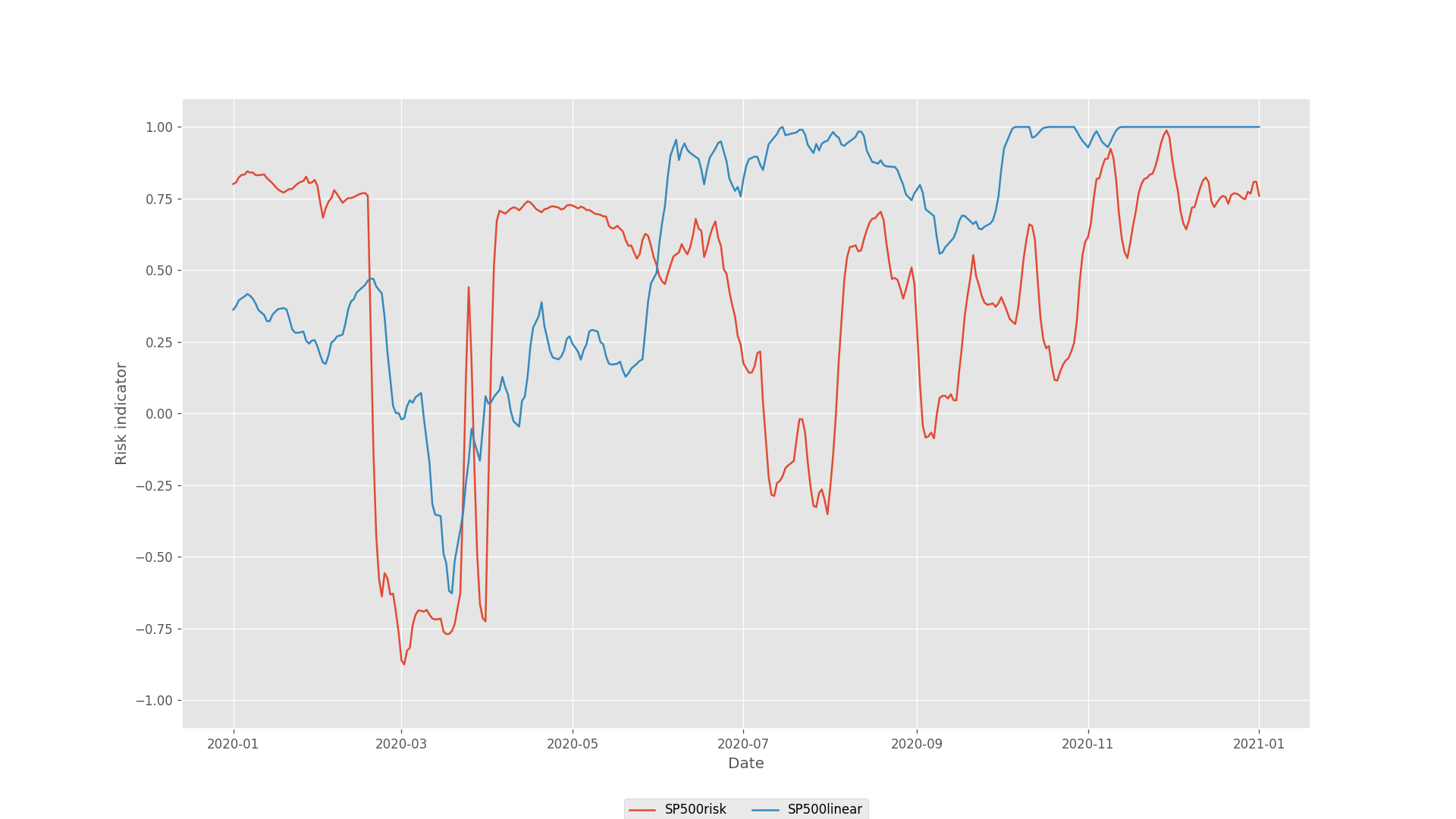}
\caption{The limits of a linear indicator: The linear risk indicator (in blue) is unable to react quickly to a market shock, as happened at the start of the 2020 pandemic. In contrast, a non-linear risk indicator (in red) is much more reactive.}
\label{fig-2020}
\end{figure*}

\subsection{The Limits of Linear Models}

Some financial institutions and asset managers resort to linear models when designing their own risk indicators. Billions of dollars in assets under management are allocated using strategies that rely on linear models.

A linear model is of the form $\hat{y} = \sum w_i x_i $, where $\hat{y}$ is the prediction (i.e. the risk indicator), the different $x_i$ are the variables, features, or factors taken into account to make the prediciton, and the weights $w_i$ model the importance of each feature. Those weights can be learnt using a standard linear regression model.

From a formal point of view, a linear model is only able to separate between linearly-separable classes. In other words, the decision frontier of such a model is a hyperplane (the generalization of a three-dimensional plane and a straight line in a two-dimensional space). A linear model cannot differentiate between non-linearly-separable classes, no matter how it is learnt. 

Given that the World is highly nonlinear, linear approximations are not always suitable. They underfit data and this underfitting causes an error that cannot be suppressed because of the intrinsic limitations of linear models. In particular, given a linear model:

\begin{itemize}
    
    \item A change $\Delta x_i$ in one of the model variables provokes a change $\Delta \hat{y} = w_i \Delta x_i$ in the model prediction.

    \item That change, $\Delta \hat{y}$ is always the same, no matter what the current context is. When $x_i$ has an associated positive weight ($w_i >0$), the prediction always changes in the same direction of the change observed in the input variable $x_i$. Likewise, a negative weight ($w_i < 0$) makes input variable and prediction change in opposite directions.

    \item As a consequence of model linearity, changes in the model output are always proportional to changes in the model inputs. In extreme situations, model predictions are slow to change, given that input changes are often gradual.
    
\end{itemize}

Figure \ref{fig-2020} illustrates the behavior of a linear model, in blue, when used to predict trends in the S\&P 500 stock index. At the end of February 2020, the effects of the COVID-19 pandemic were already affecting worldwide markets, a few weeks before WHO characterized the virus outbreak as a pandemic on March 11th, 2020. The linear risk indicator, however, was slow to react. In contrast, a nonlinear risk model, based on artificial neural networks, was much more reactive (shown in red in Figure \ref{fig-2020}). 

Artificial neural networks, currently known as deep learning models, are universal approximators from a mathematical point of view \cite{cybenko1989} \cite{funahashi1989}\cite{hornik1989}\cite{montufar2014}. Essentially, they are combinations of multiple simple mathematical functions that, when combined, can implement more complicated functions. In short, they are not subject to the stringent limitations of linear models.

When used for designing risk models, deep learning techniques are able to differentiate between non-linearly-separable classes. Their decision frontiers are no longer hyperplanes, they are virtually arbitrary.

Unlike the linear risk models used by many asset managers, ACCI relies on nonlinear risk models. Their non-linearity provides higher predictive capabilities for identifying market trends and makes them much more reactive to sudden changes in market conditions.

\subsection{On the Use of Risk Indicators by Fund Managers}

Given the complexity of financial markets, asset managers face many challenges when deciding how to allocate assets and when to change their portfolio composition. A risk indicator can help alleviate some of their burden by providing a timely signal they can use to change their risk exposure.

\begin{figure*}
\centering
\includegraphics[width=\textwidth]{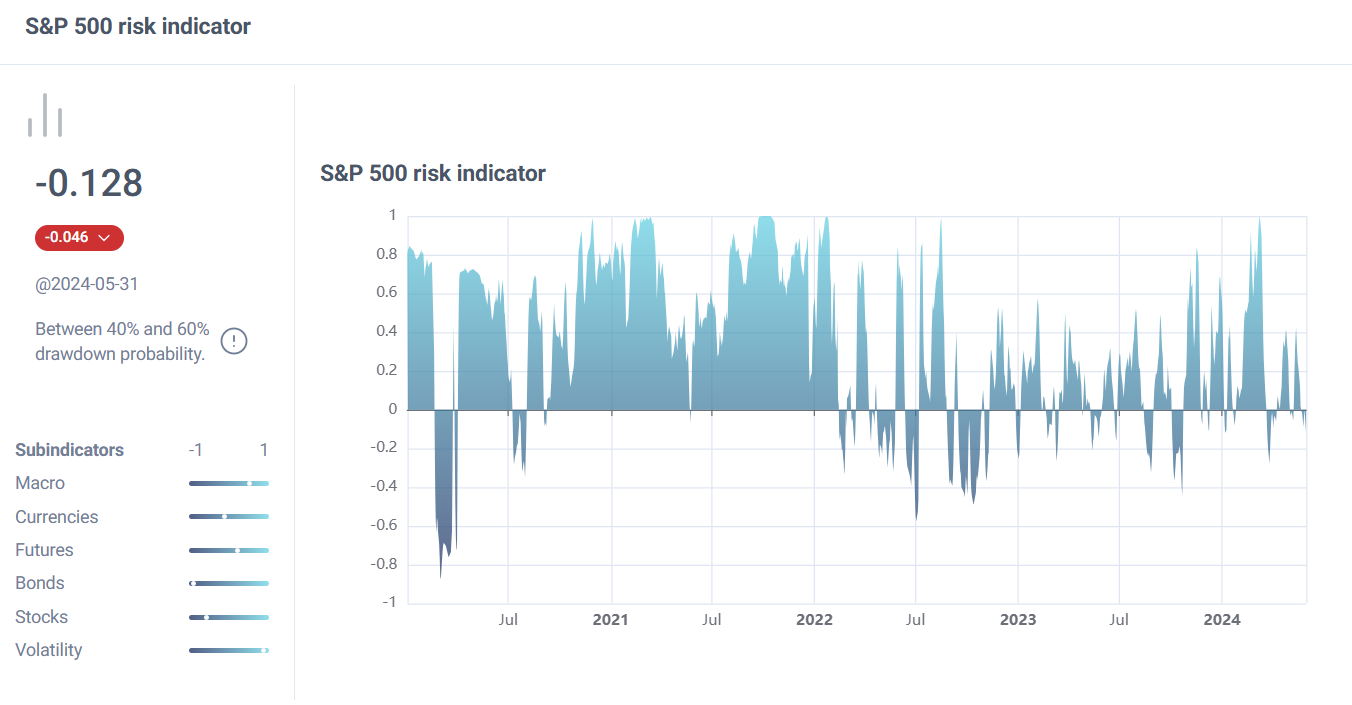}
\caption{ACCI S\&P 500 risk indicator. Source: ACCI Wealth Technologies, https://www.acciwealth.com/.}
\label{fig-sp500risk}
\end{figure*}

Figure \ref{fig-sp500risk} shows the ACCI risk indicator for the S\&P 500 stock market index. It reacted quickly at the start of the 2020 pandemic and changed its course into positive territory just a month later (the annual return of the S\&P 500 index was +16.26\% in 2020). The risk indicator remained positive in 2021 (+26.89\% S\&P 500) and often turned to a negative value in 2022 (-19.44\% S\&P 500). It remained conservative, with shorter periods of negative values, in 2023 (+24.33\% S\&P 500). Finally, in 2024, it started with a positive outlook during its first quarter (+10.79\% S\&P 500).

How can asset managers use the information provided by a risk indicator? They can track its value to modulate their risk exposure according to the current market situation. Let us recall that a risk indicator for a given market predicts the probability of a severe drawdown (e.g. $>5\%$ in equity markets, $>2\%$ in bond markets).

ACCI provides risk indicators for different markets, including the S\&P 500 stock index. In the following paragraphs, we illustrate the use of the ACCI S\&P 500 risk indicator for different scenarios that might be suitable investment strategies for particular investors:

\begin{itemize}

\item 
{\bf Risk-on/risk-off strategy} (e.g., XLK/XLP): The investor switches between two assets depending on the value of the risk indicator for the stock market. When the risk indicator is high (i.e., low drawdown probability), you invest into a procyclical sector, such as technology. In this case, we use the XLK ETF (Technology Select Sector SPDR Fund). When the risk indicator is low, i.e., below a predefined threshold (i.e., a higher drawdown probability), you opt for investing in a countercyclical sector, such as consumer staples (goods like foods and beverages, household goods, and hygiene products, as well as alcohol and tobacco, that people are unable -or unwilling— to cut out of their budgets regardless of their financial situation). In this case, we use the XLP ETF (Consumer Staples Select Sector SPDR Fund).

Figure \ref{fig-strategy-on-off} shows the 5-year returns obtained by this simple strategy, which just switches between technology stocks and consumer staples. Table \ref{tab-strategy-on-off} summarizes its performance metrics. The cumulative return of this risk-on/risk-off strategy is $192.62\%$, much better than the S\&P 500 return  ($92.30\%$), albeit its maximum drawdown ($-36.73\%$) is also higher than the maximum drawdown of its benchmark ($-33.92\%$). The portfolio volatility is similar to the overall S\&P 500 index. Sharpe and Sortino ratios are, therefore, higher due to higher returns.

\begin{table}
\begin{center}
\caption{Risk-on / risk-off performance metrics.}
\label{tab-strategy-on-off}
\begin{tabular}{lrr}
\hline
   & Portfolio & Benchmark \\
\hline
Cumulative return & 192.62\% & 92.30\% \\
Annualized return & 23.96\% & 13.97\% \\
Standard deviation & 1.321\% & 1.338\% \\
Annualized volatility & 20.96\% & 21.24\% \\
Maximum drawdown & -36.73\% & -33.92\% \\
Sharpe ratio & 2.186 & 1.617 \\
Sortino ratio & 3.035 & 2.200 \\
\hline
\end{tabular}
\end{center}
\end{table}

\begin{figure*}[!h]
\centering
\includegraphics[width= 0.86 \textwidth]{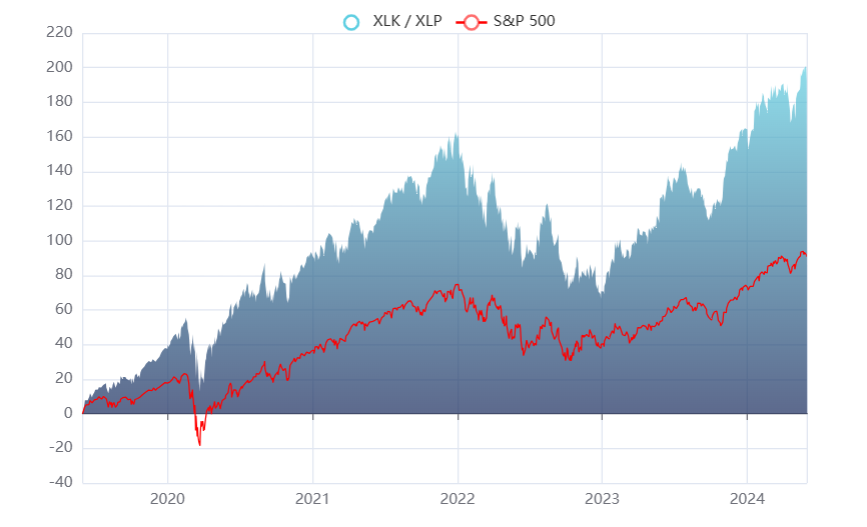}
\caption{Risk-on / risk-off portfolio cumulative returns.}
\label{fig-strategy-on-off}
\end{figure*}

\item
{\bf Cyclical strategy}, with a more diversified portfolio:
Using the same decision criteria we used in the risk-on/risk-off example, we switch between two predefined portfolios depending on the value of the risk indicator for the overall stock market. When the risk indicator is high, we use a 100\% equity portfolio with a selection of procyclical sectors, which are expected to offer positive returns when the outlook is positive. When the risk indicator is low, we reduce our equity exposure to 30\%, with a combination of countercyclical assets, suitable for more uncertain times. The composition of the procyclical (positive) and countercyclical (defensive) portfolios are are shown in Table \ref{tab-strategy-cyclical-portfolio}

Figure \ref{fig-strategy-cyclical} displays the 5-year returns obtained by our cyclical strategy, which just switches between two predefined portfolios, from 30\% to 100\% equity. Table \ref{tab-strategy-cyclical} summarizes the performance metrics associated to our cyclical/countercyclical strategy. Its cumulative return of this risk-on/risk-off strategy is $162.26\%$, still much better than the S\&P 500 return  ($92.30\%$), albeit $30\%$ lower than the risk-on/risk-off example above. Its volatility, however, is lower, given the base portfolio diversification. Sharpe and Sortino ratios are now higher than before due to higher returns and lower volatility.

\begin{table}
\begin{center}
\caption{Cyclical/countercyclical performance metrics.}
\label{tab-strategy-cyclical}
\begin{tabular}{lrr}
\hline
   & Portfolio & Benchmark \\
\hline
Cumulative return & 162.26\% & 92.30\% \\
Annualized return & 21.27\% & 13.97\% \\
Standard deviation & 1.019\% & 1.338\% \\
Annualized volatility & 16.17\% & 21.24\% \\
Maximum drawdown & -31.96\% & -33.92\% \\
Sharpe ratio & 2.433 & 1.617 \\
Sortino ratio & 3.369 & 2.200 \\
\hline
\end{tabular}
\end{center}
\end{table}

\begin{figure*}[!h]
\centering
\includegraphics[width=0.86 \textwidth]{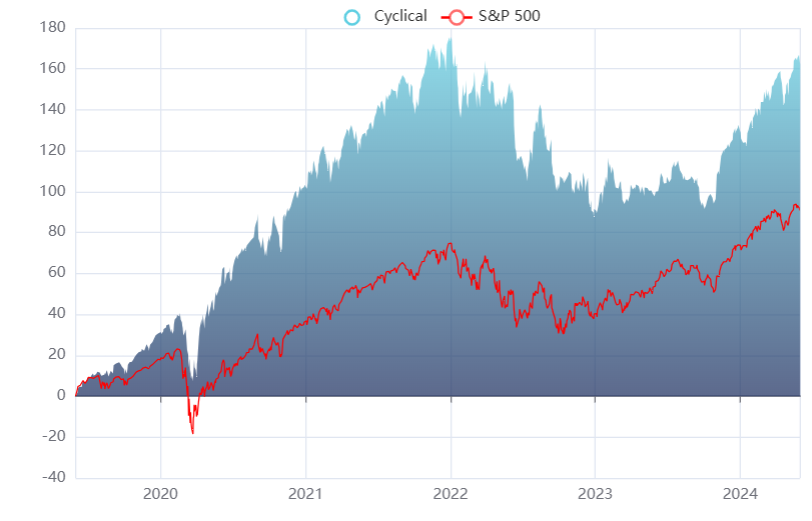}
\caption{Cyclical portfolio cumulative returns.}
\label{fig-strategy-cyclical}
\end{figure*}

\begin{table}
\begin{center}
\caption{Cyclical/countercyclical portfolios.}
\label{tab-strategy-cyclical-portfolio}

\begin{tabular}{rl}
     & Defensive portfolio (30\% equity) \\
\hline
20\% & US Bond 0-1yr iShares ETF (Acc) \\ 
20\% & US Bond 1-3yr iShares ETF (Acc) \\ 
10\% & Gold Futures \\ 
10\% & Oil Futures (Brent) \\ 
10\% & S\&P-GSCI Commodity Index Future \\ 
10\% & XLE Energy SPDR Select Sector ETF \\ 
10\% & XLU Utilities SPDR Select Sector ETF \\ 
10\% & XLP Consumer Staples SPDR Select Sector ETF \\
\hline
     & \\
     & Positive portfolio (100\% equity) \\
\hline
20\% & S\&P 500 2x leveraged ETF \\ 
20\% & S\&P 500 ETF \\ 
20\% & NASDAQ ETF \\ 
10\% & XLK Technology SPDR Select Sector ETF \\ 
10\% & XLY Consumer Discretionary SPDR Select Sector ETF \\ 
10\% & SOXX Semiconductor ETF \\
10\% & IBB Biotechnology ETF \\
\hline
\end{tabular}
\end{center}
\end{table}

\item
{\bf Systematic allocation strategy}. 
As a third example, we include a complete strategy with 3 different base portfolios adjusted to different risk levels. These base portfolios cover a more diversified asset base and comply with the requirements of the UCITS [Undertakings for Collective Investment in Transferable Securities] regulatory framework in the European Union. The actual ACCI SA fund now employs a similar asset allocation strategy.

For negative risk indicator values, we use a 30\% equity defensive portfolio, e.g. 30\% in the S\&P index and the remaining 70\% in (mostly short-term) US bonds. For intermediate risk indicator values, we define a 70\% equity balanced portfolio, with the equity part mostly in the S\&P index and partially in emerging markets, whereas we keep 20\% in short-term US bonds and 10\% in money markets. Finally, for positive risk indicator values, we have a 100\% equity positive portfolio, now including NASDAQ ETFs. The design of these base portfolios is gradual, so that changing from one to the next does not involve a 100\% portfolio rotation. In our example, from the defensive to the balanced portfolio, we would just exchange 40\% of our portfolio assets (keeping 30\% in fixed-income assets and 30\% in equity markets), to meet our 70\% equity requirements. Likewise, the balanced and positive portfolios can be designed to minimize portfolio rotation when switching from the former to the latter or vice versa.

Figure \ref{fig-strategy-sa} displays the 5-year returns obtained by our systematic allocation strategy, which switches among three predefined  portfolios (30\%/70\%/100\% equity). Table \ref{tab-strategy-sa} summarizes the performance metrics associated to our SA strategy. Its cumulative return of this risk-on/risk-off strategy is still better than the S\&P 500 return, $127.58\%$, and much better than its associated benchmark (50\% MSCI World + 50\% Global Aggregate). Its volatility is kept under control and it exhibits a reduced maximum drawdown. Sharpe and Sortino ratios are even higher than in the previous examples.

\begin{table}
\begin{center}
\caption{Systematic allocation performance metrics.}
\label{tab-strategy-sa}
\begin{tabular}{lrr}
\hline
   & Portfolio & Benchmark \\
\hline
Cumulative return & 127.58\% & 22.24\% \\
Annualized return & 17.88\% & 4.10\% \\
Standard deviation & 0.659\% & 0.642\% \\
Annualized volatility & 10.47\% & 10.20\% \\
Maximum drawdown & -17.19\% & -24.70\% \\
Sharpe ratio & 3.059 & 0.982 \\
Sortino ratio & 4.381 & 1.332 \\
\hline
\end{tabular}
\end{center}
\end{table}

\begin{figure*}[!h]
\centering
\includegraphics[width=0.86 \textwidth]{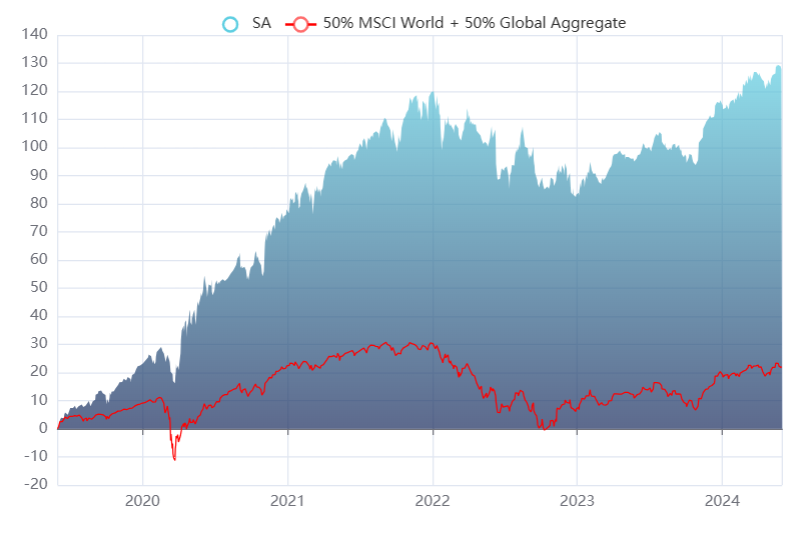}
\caption{Complete systematic allocation portfolio returns, subject to UCITS constraints.}
\label{fig-strategy-sa}
\end{figure*}

\end{itemize}

As shown above, risk indicators can serve as a guideline to configure different investment strategies. From switching between assets, as in our risk-on/risk-of example, to full-fledged portfolio allocation, as illustrated by our systematic allocation case study.

The use of risk indicators can be tailored to the preferences of a particular asset manager. He can just set a threshold to decide when to switch from a defensive portfolio to a positive one, or vice versa. Or he can adjust his risk exposure daily, according to the risk indicator value at the close of the previous market session. 

In the examples above, we assumed that the threshold was a predefined value, the same for our 5-year simulations. However, thresholds can be adjusted dynamically, in accordance to the manager outlook for a given time frame. 

Hysteresis can also be used to minimize the number of buy/sell operations. For instance, if we assume a 10\% margin, which corresponds to 0.2 points in the $[-1,+1]$ interval of the risk indicator, we can switch to our defensive portfolio when the risk indicator falls below 0.0, yet do not return to our positive portfolio until the risk indicator raises over 0.2. In that case, minor fluctuations in the risk indicator are not translated into potentially unnecessary flip-flop rotations in our portfolio composition.

In summary, risk indicators can provide an important signal for asset management, yet keeping the human asset manager fully in charge of the situation. He can adopt their use in the way that best serves his purposes. Given his particular goals, he can customize his investment strategy by integrating the use of risk indicators within his own comfort zone.

\section{Model Training for Market Trend Prediction}

In the previous section, we introduced the use of risk indicators in asset management. In this section, we delve into the details of how they can be designed for particular markets. 

As described above, risk indicators are predictive models for drawdown periods. Drawdown, when talking about investments, is a measure of the decline from a previous historical peak in the cumulative return or current value of an investment strategy. In other words, we focus on the downside risk, the probability that an asset portfolio will fall in price. Given historical data, Machine Learning techniques can be used to model that probability.

\subsection{Machine Learning}

\begin{figure}[t]
\centering
\includegraphics[width=3in]{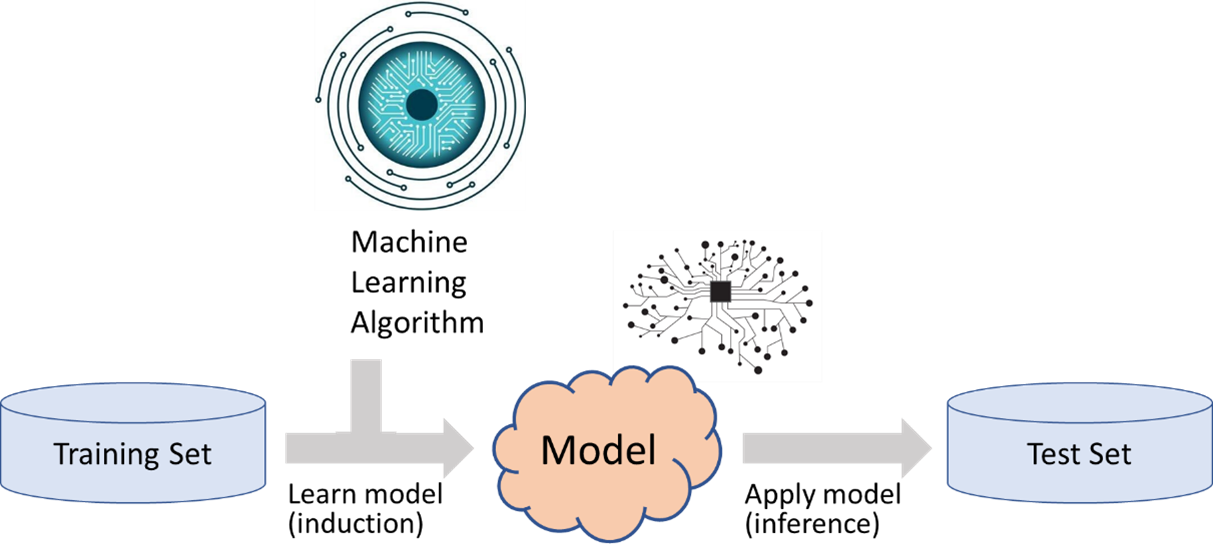}
\caption{Machine Learning: Learning from data using Artificial Intelligence.}
\label{fig-ml}
\end{figure}

Machine Learning [ML] is a field within Artificial Intelligence [AI] that studies the design and development of algorithms that can learn from data. Hopefully, the models learnt using ML, apart from working properly with the data they were trained on, should also generalize well to unseen data. Figure \ref{fig-ml} displays how ML works. 

Given a data set, known as training set, and a ML algorithm, the computer learns a model from the provided data. By means of an inductive process, which depends on the particular learning algorithm chosen, we build a model, whose properties also depend on the specifics of the ML algorithm. Formally, induction is making an inference based on an observation of a sample (i.e., the training set). Abduction is making a probable conclusion from what you know, so model building or training, as it is often called, is an example of abductive reasoning from a mathematical logic point of view. Model training, as abductive reasoning, seeks the simplest and most likely conclusion from a set of observations (those in the training set).

Once the model is trained, it can be applied on new data to make predictions. This application of the trained model to data is an example of deductive reasoning. The model is used to draw valid inferences that follow logically from their premises (i.e., those represented by the trained model). Hence, the use of ML models is often referred to as `inference.' 

When evaluating ML models, a separate dataset is kept, distinct from the training dataset. This dataset, often called test set, is employed to evaluate the performance of ML models. Why? Because the results on the training dataset would be utterly optimistic and we need an unbiased estimation of the true performance of ML models before they are deployed. This is the role of the test set.

\subsection{Machine Learning Techniques}

We use Machine Learning techniques to design risk indicators for financial markets, both fixed-income and equity markets. These risk indicators are predictive models for drawdown periods. Since they use labelled historical data to be trained on, they can be built using supervised ML techniques.

In supervised learning techniques, training data contains both input variables (e.g., a vector of predictor variables) and the desired model output. From input-output pairs, a model is learnt from the training dataset. Where do the outputs come from? Typically, an human expert has labelled each training set example with the desired output. Once the training set is prepared, it is the turn of the computer to learn from it and train a suitable model using a particular learning algorithm.

A wide range of supervised learning algorithms are available, each one with its own strengths and weaknesses. In fact, a well-known theoretical result, the ``no free lunch theorem,'' asserts that there is no single learning algorithm that works best on all supervised learning problems \cite{wolpert1996}, a result that also applies to optimization techniques \cite{wolpert1997}.

Given that we cannot know beforehand which particular ML technique will work best for a particular problem (we might have some hints, yet they are never conclusive), we have tested multiple ML techniques in the ACCI·ON project to design ACCI risk indicators. 

Testing multiple ML algorithms lets us compare the differences among the ML models they train. Our comparison takes into account both quantitative and qualitative aspects. From a quantitative point of view, we are interested in model accuracy, precision, and recall, as well as in the episodic behavior of the risk indicator when market trends change. For us, the dynamic response of a risk indicator to a trend reversal is paramount, as we discussed when describing the limitations of linear models. From a qualitative point of view, model interpretability is desirable, yet not essential, but model behavior is crucial. Even when a binary output model might yield better quantitative results, a zero-one response does not help asset managers feel confident on the risk indicator value, as changes in the indicator seem to be unpredictable. A gradual response is often preferable, when daily changes in the risk indicator hint at current potential trends in the underlying market.

In the ACCI·ON project, a wide range of supervised ML techniques have been evaluated:

\begin{itemize}


\item 
{\bf Linear models}, even when they exhibit some undesirable properties, such as their lack of responsiveness when a market trend is reversed, are still useful as a baseline. They provide a foundation on which we can build on to compare the effectiveness of more sophisticated learning algorithms. In our experiments, we tested linear regression for regression problems as well as logistic regression for classification problems.


\item
{\bf Support vector machines}: A support vector machine, or SVM, in addition to linear classification, can efficiently perform a non-linear classification using what is called the kernel trick. Since linear approximations are not suitable for the non-linear world of financial markets, SVMs provide an interesting alternative, even though they are not truly scalable. SVMs represent data through pairwise similarity comparisons between original data observations. SVMs implicitly represent the original data in transformed coordinates within a higher dimensional space (actually, a potentially infinite-dimensional space) and identify the maximum-margin hyperplace in that space. Even though the decision frontier is still linear in the transformed coordinate space, it corresponds to a non-linear frontier in the original space. SVMs can be used to solve both classification problems \cite{cortes1995} and regression problems \cite{drucker1997}.



\item 
{\bf Ensembles} are popular for winning data mining competitions. In ML, ensemble methods combine multiple learning algorithms. As musical ensembles combine multiple musical instruments to achieve a more harmonious result, ML ensembles obtain better predictive performance than any of the individual learning algorithms in the ensemble. For that to occur, the ensemble must be designed so that we can ensure that the individual algorithms within the esemble do not always make the same mistakes. From a quantitative point of view, they can achieve the best numerical results, hence their popularity in Kaggle competitions \cite{bojer2021}, even though they might be unsuitable from a qualitative point of view. Two of the best-known ensemble learning algorithms are random forests \cite{breiman2001} and gradient boosting \cite{friedman2001}. 

\begin{itemize}

\item 
A random forest, proposed by Leo Breiman from the University of California, Berkeley, is an ensemble of decision trees. A decision tree is a symbolic model most economists are already familiar with. Decision trees were very popular in Machine Learning and Data Mining at the turn of the century.

\item 
Gradient boosting, proposed by Jerome Friedman from Stanford University, is based on boosting. Most boosting algorithms consist of iteratively learning weak classifiers and adding them to a final strong classifier. A weak classifier is only slightly correlated with the true classification (it can label examples better than random guessing). The resulting strong learner is a classifier that is arbitrarily well-correlated with the true classification. The predition model obtained by gradient boosting is an ensemble of weak prediction models. Gradient boosting algorithms are iterative functional gradient descent algorithms; that is, they optimize a cost function over function space by iteratively choosing a function (weak hypothesis) that points in the negative gradient direction.

\end{itemize}


\item
{\bf Deep learning} models are based on artificial neural networks \cite{goodfellow2016} \cite{berzal2019a} \cite{berzal2019b}. Artificial neural networks are connectionist models, formerly known as Parallel Distributed Processing (PDP) models.  

Neural networks consist of individual neurons, which are simple computational elements of the form
$$ y = f \left( \sum w_i x_i\right) = f(\vec{w} \cdot \vec{x})$$

The nonlinear function $f$ is the neuron activation function, typically a sigmoidal function, such as the logistic function and the hyperbolic tangent, or a rectified linear function. In the former case, the neuron is said to be sigmoidal; in the latter, it is a ReLU [Rectified Linear Unit].

Multiple neurons can be put in parallel to create a network layer with vector input $\vec{x}$, vector output $\vec{y}$, weight matrix $W$, and activation function $f$, to be applied element-wise:
$$ \vec{y} = f \left( W \vec{x} \right)$$

Multiple network layers can be stacked to create a feed-forward neural network:
$$ \vec{y} = f \left( W_L f \left( W_{L-1} ... f \left( W_1 \vec{x} \right)  \right) \right)$$

The last layer, characterized by the weight matrix $W_L$, is the network output layer. The input vector $\vec{x}$ is the input layer, which performs no computation, just provides the input to the network. Inner layers are called hidden layers. When the network has more than one hidden layer, the network is said to be a deep neural network, hence the term `deep learning' to refer to the learning techniques that allow us to train deep neural networks.

The weights of an artificial neural network are typically trained by stochastic gradient descent with the help of a dynamic programming algorithm called backpropagation. Backpropagation is an efficient gradient estimation method for neural network models, also known as the reverse mode of automatic differentiation or reverse accumulation. Backpropagation computes the gradient of a loss function with respect to the weights of the network for a single input–output example. It computes the gradient one layer at a time and iterates backward from the last layer to avoid redundant calculations of intermediate terms in the Leibniz chain rule that is applied to compute the gradient.

In contrast to symbolic models (e.g., decision trees) and statistical techniques (e.g. SVMs), neural networks were originally proposed as computational models to describe aspects of human perception, cognition, and behaviour, the learning processes underlying such behaviour, and the storage and retrieval of information from memory \cite{mcclelland2009}. 

From a computational perspective, feed-forward neural networks can be interpreted as models that learn to extract hierarchical features from data. The first layers of a feed-forward network learn to extract relatively simple features directly from the input data. As we advance through a deep network, neurons learn to represent more complex features from the features extracted by previous network layers. Deep learning can, therefore, be viewed as hierarchical feature representation \cite{bengio2009}, hence the name of one of the major conferences in the area (ICLR, the International Conference on Learning Representations).
\end{itemize}

Training ML models from real-world data poses significant practical challenges. The design of the proper predictive model for a given problem involves making decisions with respect to multiple aspects of the ML model. We cover the different degrees of freedom we have when building predictive models for market trend prediction in the following subsections.

\subsection{Model Output}

The first decision we must make when building a predictive model is the nature of our target variable, the value we are trying to predict, typically denoted by $\hat{y}$. If we choose to predict a categorical, discrete, or nominal variable, we build a classification model. If we opt for predicting a continuous real-valued variable, we are building a regression model.

Market trend prediction can be modeled either as a classification problem or as a regression problem:

\begin{itemize}

\item 
{\bf Trend prediction as a classification problem}:

Our target variable will be a binary variable that indicates whether or not our market of interest is immersed in a drawdown period. The drawdown period covers from peak to trough, from a local maximum to the following local minimum.

How are local maxima and minima chosen? There are several alternatives:

\begin{itemize}

\item
We can choose the top-k drawdown periods according to their magnitude. In the S\&P 500 index, the largest market crashes and bear markets include the crash of 1929 (-86\% from September 1929 to June 1932), the bear market of 1937 (-54\% from March 1937 to March 1938), the 1973 oil shock (-48\% from January 1973 to October 1974), the 1987 bear market (-34\% from August 1987 to December 1987, including the Black Monday of October 19, 1987), the burst of the dot-com bubble (-49\% from March 2000 to October 2002), the Global Financial Crisis (-48\% from August 2008 to March 2009), and the start of the Covid pandemic (-33.92\% from February 19th to March 23rd, 2020).

\item 
Alternatively, we can choose every drawdown period where our market of interest drops beyond a predefined threshold (in percentage points). For instance, a fund manager might be interested in predicting equity drawdowns above 5\%, deeming it unnecessary to act when downward trends are smaller in magnitude, yet a different asset manager might be interested in predicting 2\% drops in bond markets.
\end{itemize}

In both cases, we might establish a time horizon, the maximum period of time to be considered part of the current trend. This time horizon, depending on the situation, might be measured in days, weeks, months, or quarters, even years.

For classification problems, the cross-entropy, also known as logarithmic loss or log loss, is a suitable loss function for training a predictive model:
$$ CE = - [ y \log \hat{y} + (1-y) \log (1 - \hat{y}) ]$$
where $y$ is the desired output and $\hat{y}$ is our prediction. 

\item 
{\bf Trend prediction as a regression problem}:

We can also interpret our trend prediction goal as a regression problem. In this case, we can try to predict:

\begin{itemize}

\item
The magnitude of the expected market drawdown at each time period: 0 in bull markets, the maximum drawdown at the start of a bear market, and 0 at the end of the drawdown period corresponding to the bear market.

\item
The expected market return until the next market trend reversal: positive and decreasing in bull markets; negative and increasing in bear markets.

\item
The overall market return during the whole current market trend: positive and constant in bull markets, negative and constant in bear markets.

\item
The overall market drawdown during a bear market: zero in bull markets, negative and constant in bear markets.

\end{itemize}

As before, we might consider an unlimited time horizon (for bear markets) or set a predefined time horizon in accordance to the frequency of our particular trading strategy.

For regression problems, the mean squared error is a suitable loss function for training our predictive models:
$$ MSE = \frac{1}{N} \sum_{i=1}^{N} \left ( y - \hat{y} \right)^2$$
where $y$ is the desired output and $\hat{y}$ is our prediction.

\end{itemize}

No matter if we model our prediction as a classification or regression problem, we must also take into account that, for our prediction to be actionable, it must be of the form $\hat{y}(t+2) = f(x(t))$. Our prediction at the close of a trading session is a prediction valid not for the immediately-following trading session, but for the session after that. In other words, we must have the opportunity to trade today using yesterday's session data to prepare for tomorrow's trend.

\subsection{Input Variables}

Simple forecasting models are autoregressive. In statistics, econometrics, and signal processing, the output or target variable of an autoregressive model depends only on its own previous values. In time-series analysis, we  predict the future values of a time series based on its past values. Autoregressive models are widely used in technical analysis to forecast future security prices.

More advanced forecasting models take additional context into account. That context can be provided by additional variables or time series. For instance, instead of predicting future S\&P 500 values using only past S\&P 500 values, we might also incorporate bond market data as an additional input to our predictive model.

With the initial support of Umberto Mármol and other economists at ACCI Capital Partners, we analyzed a multitude of economic variables that might serve as leading indicators for predicting changes in market trends. It is essential that they are leading indicators because we are especially interested in detecting changing trends as soon as they happen. Many economic variables are either lagging indicators (they hint at trends after they have started) or are published with too much delay to be useful when we expect a quick response from our trend prediction model. These were finally discarded in our risk indicator models.

Our final risk indicator models incorporate dozens of different variables, sometimes hundreds. In broad terms, the variables we use as input can be grouped into the following six categories:

\begin{itemize}

\item 
Stock market indexes for main global and regional equity markets, including the S\&P 500, the MSCI World, the NASDAQ 100, or the Russell 2000, as well as the S\&P 500 Equal Weight Index and many other regional stock market indexes.

\item
Bond market data, including US Treasury bonds, their yield curve, government bonds from the major economies of the World, commercial paper interest rates, and corporate bonds (both investment-grade and high-yield).

\item 
Currency exchange rates for the World major currencies and currency baskets such as the U.S. Dollar Index (DXY).

\item
Futures market data, including commodity indexes (GSCI and DJCI), commodity futures, energy (oil and natural gas), and precious metals (e.g., gold).

\item
Volatility indexes associated to different markets, including the well-known VIX and MOVE volatility indexes, as well as volatility measures for commodities, gold, currencies, stocks, and bonds.

\item 
Macroeconomic variables including leading indexes (OECD and BBK), ISM data, freight indexes, advance retail sales, consumer sentiment data, monetary data, or weekly initial unemployment claims, among many others.
\end{itemize}

A wide range of input variables were chosen for their ability to move the particular markets we were designing risk indicators for or just for their usefulness as leading indicators. Using scores, even hundreds, of input variables in a predictive model causes technical problems that are not always solvable using traditional techniques. Fortunately, deep learning models were up to the task and helped us improve the predictive power of our risk indicators.

\subsection{Feature Engineering}

Once input variables have been selected, they must be prepared to be used as the input to Machine Learning models. First of all, a proper encoding must be selected, often depending on the particular ML technique to be used for training a predictive model.


Even when no additional variables are included, apart from those of the time series we wish to model, we must decide whether we provide the time series values as they are acquired or we preprocess them to make it easier for the ML algorithm to learn a good model. For instance, when we are predicting the evolution of the S\&P 500 stock index, using index values would hamper most ML algorithms, since the index is near its all-time high and current index values have never been observed in the past. It is much more reasonable to use percentage changes, always as model input and as model output in the case of regression models.


Before using some ML techniques, the scale of the input data must also be adjusted. Even when ML techniques are able to cope with different scales for different inputs, learning is easier if we do some preprocessing. It is usually a good idea to normalize or standardize all the model inputs before proceeding further. Scale changing transformations include the following:

\begin{itemize}

\item 
Feature scaling, unity-based normalization, or [0,1] normalization brings all values into the [0,1] unit interval:
$$ x_{[0,1]} = \frac{x - x_{min}}{x_{max}-x_{min}}$$

\item
Min-max feature scaling or min-max normalization brings values into the [a,b] interval:
$$ x_{[a,b]} = a + \frac{x - x_{min}}{x_{max}-x_{min}} \left( b - a \right)$$
For instance, some deep learning models benefit from a bipolar encoding, using the [-1,1] interval.

\item
Robust normalization or robust standardization employs the median and interquartile range (IQR) to be more robust against outliers in data:
$$ x_{robust} = \frac{x - x_{median}}{x_{IQR}}$$

\item
Standardization or z-score normalization is a more common approach, using the mean $\mu$ and the standard deviation $\sigma$:
$$ z = \frac{x - \mu}{\sigma}$$
The z-score, or standard score, measures the number of deviations by which the value of the raw score is above or below the mean value. It works well for data that is normally distributed, when large deviations from the mean are not frequent. In a normal distribution, 68.3\% of the data lie in the $[\mu-\sigma, \mu + \sigma]$ interval, 95.4\% of the data lie in within two standard deviations (the $[\mu-2\sigma, \mu + 2\sigma]$ interval), and  99.7\% of the data lie in within three standard deviations (the $[\mu-3\sigma, \mu + 3\sigma]$ interval). The six-sigma interval covers 99.9999998\% of data (less than one value in 500 million falls outside this range).

When you have a sample of data, the z-score computation of that sample yields the well-known t-statistic, used by Student's t-test for statistical hypothesis testing.
\end{itemize}


When working with time series, not all ML techniques are able to use them directly as model inputs. Therefore, scalar features are often extracted from them. Autoregressive models are often limited to a small window in the time series past; i.e., they use just the previous time series values as input. Some deep learning models, fortunately, are able to process sequences in general and time series in particular.


The efficient market hypothesis [EMH] states that prices reflect all available information and consistent alpha generation is impossible. If it were true, beating the market would not be feasible. However, even simple autoregressive models can benefit from the design of input variables derived from the original time series we are modeling. For instance, Zura Kakushadze, from Quantigic Solutions LLC, Stamford, CT \cite{kakushadze2016}, enumerates 101 `alphas' that have proven to be useful in algorithmic trading. An `alpha' is an indicator that, when used in combination with historical data, can be useful for making predictions on the future price movements of financial instruments. In other words, `alphas' provide some capability to beat the market. When used for predicting the S\&P 500 stock index, for instance, they achieve a 54\% accuracy rate \cite{fernandez2022}, somewhat above the 50\% accuracy expected by the EMH, an edge that is enough to obtain benefits if we used high frequency trading [HFT] strategies.

\begin{figure}[!t]
\centering
\includegraphics[width=3in]{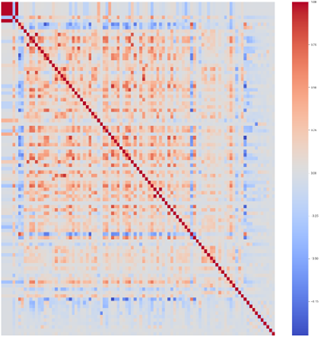}
\caption{Correlation matrix for 101 alphas, from \cite{fernandez2022}: Indicators employed by algorithmic trading strategies often exhibit some correlation between them.}
\label{fig-alphas}
\end{figure}


High frequency trading, however, is not a suitable investment strategy for many investment funds. More traditional asset managers do not want to trade too often, so the quantitative models used in HFT do not match their expectations. They often want to minimize the number of trades in their portfolios (in order to reduce operational risks) and prefer a low portfolio rotation. Risk indicators, as described in this paper, are designed for them.

There is still a problem that must be solved. Daily market data is noisy. That noise might cause sudden temporary fluctuations in risk indicators, leading to unnecessary trades and flip-flop rotations. Depending on the particular ML technique employed to build the risk indicator, different noise reduction strategies might be used:

\begin{itemize}

\item
Some ML algorithms, such as linear models, are not particularly robust to the presence of noise in data, so that noise must be filtered before it reaches the model input. Input noise filtering, however, delays the input to the model when trend changes suddenly appear. This delay is added to the limited reactivity of linear models, hence compounding the problem and limiting the ability of linear risk indicators to react to trend reversals.

\item
Other ML algorithms, such as deep learning models, can be made robust against input noise without having to filter that noise beforehand. In fact, noise can act as as regularizer and help improve the performance of such models on unseen data. If brief fluctuations are observed in their output, that output can be filtered to reduce wiggling and provide more aesthetically pleasing risk indicators, with a more continuous appearance and without the loss of model responsiveness associated to input noise filtering.

\end{itemize}


Apart from data normalization/standardization, custom feature engineering, and noise reduction filters, automated feature extraction and dimensionality reduction techniques can also help improve the performance of trend prediction models. Some of alternatives available for mining complex data include the following:

\begin{itemize}

\item 
{\bf Principal component analysis [PCA]} is the best-known feature extraction and dimensionality reduction technique, developed by Pearson in 1901 \cite{pearson1901} and Hotelling in 1933 \cite{hotelling1933}. Data is linearly transformed onto a new coordinate system so that that the new axes (i.e. the principal components) capture the largest variation in the data can be easily identified. In other words, the coordinate system is rotated to match the distribution of the input data and most of the variance in data is captured by the first few dimensions.

\item 
{\bf CUR decomposition} \cite{drineas2006} \cite{mahoney2006} \cite{mahoney2008} approximately expresses the original data in terms of a basis consisting of actual data elements and thus have a natural interpretation in terms of the processes generating the data. CUR decomposition is a scalable alternative to PCA, when data can be stored but is too large to practically perform superlinear polynomial time computations on it.

\item 
{\bf Dynamic Mode Decomposition [DMD]} \cite{schmid2008} \cite{schmid2010}, given time series data, computes a set of modes each of which is associated with a fixed oscillation frequency and decay/growth rate. Each mode can be physically meaningful because it is associated with a damped (or driven) sinusoidal behavior in time.

\item
{\bf Wavelets}, sometimes called brief oscillations, are wave-like oscillations with an amplitude that begins at zero, increases or decreases, and then returns to zero one or more times. Used for decades in digital signal processing, the first-known wavelet was the Haar wavelet (1909). Wavelet analysis is similar to Fourier analysis in that it allows a target function over an interval to be represented in terms of an orthonormal basis. A key difference is that wavelets capture both frequency and location information (i.e., location in time).

\item
{\bf Kernel PCA} \cite{scholkopf11998} is is a nonlinear extension of principal component analysis (PCA) using kernel methods. The kernel trick is used to factor away much of the computation: a non-trivial, arbitrary function is 'chosen' that is never calculated explicitly, allowing the possibility to use very-high-dimensional 
functions, since we never have to actually evaluate the data in that space.

\item
{\bf Stochastic Neighbor Embedding [SNE]} \cite{hinton2002} places objects, described by high-dimensional vectors or by pairwise dissimilarities, in a
low-dimensional space in a way that preserves neighbor identities. Its more efficient t-distributed variant, t-SNE \cite{vandermaaten2008}, runs in $O(n^2)$ time and requires $O(n^2)$ space.

\item
{\bf Uniform manifold approximation and projection [UMAP]} \cite{mcinnes2018} is another nonlinear dimensionality reduction technique. Visually, it is similar to t-SNE, but it assumes that the data is uniformly distributed on a locally connected Riemannian manifold.

\item
{\bf Autoencoders} are deep learning models used to learn efficient codings of unlabeled data (i.e., for unsupervised learning). In fact, they were originally proposed as a nonlinear variant of PCA based on artificial neural networks \cite{kramer1991}. An autoencoder learns two functions: an encoding function that transforms the input data, and a decoding function that recreates the input data from the encoded representation. Regularized autoencoders (sparse \cite{makhzani2013}, denoising \cite{vincent2010}, and contractive \cite{rifai2011}) are effective in learning representations for subsequent predictive tasks.
\end{itemize}

Even though you might think that given the desired model output and the designed model inputs is enough to let the computer do its job and learn the best possible model from the available data, there are still other degrees of freedom in the design of predictive models.

\subsection{Model Hyperparameters}

Each Machine Learning technique is designed for building a particular kind of model (e.g., a symbolic decision tree, a support vector machine, a neural network, or a full ensemble of models). Machine learning train those models by learning from the training data. However, there are many other model parameters, commonly known as hyperparameters, that are not learnt from data. They are adjusted by data scientists who build models for solving a particular problem.

In the case of deep learning models, the problem is daunting. You can choose different neural network topologies, i.e., different ways to connect neurons within a neural network. You can also play with the number of layers in the network or the number of neurons at each network layer. You can even change the kind of neurons that constitute the network, e.g., changing their internal structure or their nonlinear activation functions. But, beyond the neural network architecture itself, you have multiple degrees of freedom for training neural networks. There are multiple optimization algorithms and training strategies, from stochastic gradient descent (as in online or mini-batch learning) to Hessian-free optimization, using conjugate gradients or L-BFGS. There are also a wide range of heuristic techniques, known as regularization methods, that help prevent overfitting. 


Let us start by analyzing the different neural network architectures that can be used for working with time series. Time series are challenging for many ML algorithms, a problem that is exacerbated when we have to deal with hundreds of them. Fortunately, different neural networks architectures have been proposed to deal with sequential data, which includes time series as a particular case:

\begin{itemize}

\item 
{\bf Recurrent neural networks [RNNs]}:
In feed-forward neural networks, the outputs of each layer serve as inputs to the following layer, but there are no feedback loops. In recurrent neural networks, there are recurrent connections between hidden units. These recurrent connections provide feedback loops and, in some sense, provide memory to neural networks. RNNs can then operate on sequences of inputs $x_t$, with the time step $t$ representing the position of a particular within the sequence or time series.

Simple recurrent networks [SRNs] are fully-recurrent neural networks that connect the output of all hidden units to their input for each hidden layer in the network \cite{elman1990}:
$$ h_t = f(W x_t + U h_{t-1})$$
The output of hidden layers now depends on both their current input $x_t$, through their weight matrix $W$, and their previous output $h_{t-1}$, through a second weight matrix $U$.

Training RNNs is performed by backpropagation through time, ot BPTT \cite{mozer1989}, which consists of unrolling or unfolding the recurrent network and performing backpropagation on the unrolled computational graph associated to the network.

Gated RNNs create paths through time whose derivatives do not vanish nor explode, a common problem with deep neural networks and Elman's RNNs. 

The long short-term memory [LSTM] model introduced self-loops to enable paths where the gradient can flow for long \cite{hochreiter1997}. The behavior of a LSTM cell with a forget gate is defined by the following equations:
\begin{align*}
f_t &= \sigma_g(W_{f} x_t + U_{f} h_{t-1} + b_f) \\
i_t &= \sigma_g(W_{i} x_t + U_{i} h_{t-1} + b_i) \\
o_t &= \sigma_g(W_{o} x_t + U_{o} h_{t-1} + b_o) \\
\tilde{c}_t &= \sigma_c(W_{c} x_t + U_{c} h_{t-1} + b_c) \\
c_t &= f_t \odot c_{t-1} + i_t \odot \tilde{c}_t \\
h_t &= o_t \odot \sigma_h(c_t)
\end{align*}
where, as in simple RNNs, the matrices $W$ and $U$ contain the weights for inputs and recurrent connections. A LSTM cell contains an input gate $i$ that decides which pieces of new information to store in the current state, an output gate $o$ that controls which pieces of information in the current state to output, a forget gate $f$ that decides what information to discard from a previous state b, a hidden state vector $h$ (i.e., the output of the LSTM cell), and a cell state vector $c$.

Gated recurrent units [GRUs] \cite{cho2014} are like LSTMs, with a gating mechanism to input or forget certain features, but without a context vector or output gate, resulting in fewer parameters than LSTMs:
\begin{align*}
z_t &= \sigma(W_{z} x_t + U_{z} h_{t-1} + b_z) \\
r_t &= \sigma(W_{r} x_t + U_{r} h_{t-1} + b_r) \\
\hat{h}_t &= \phi(W_{h} x_t + U_{h} (r_t \odot h_{t-1}) + b_h) \\
h_t &=   (1-z_t) \odot  h_{t-1} + z_t \odot  \hat{h}_t
\end{align*}
where $\odot$ represents the Hadamard product (i.e., element-wise multiplication), $z$ is the update gate, $r$ is the reset gate, and $h$ is the output vector.

The recurrent networks we have discussed have a causal structure, their state at time $t$ only depends on data from the past, which is suitable for time series prediction. Other applications, however, employ bidirectional RNNs and their prediction depends on whole input sequences. Such architecture is suitable for dealing with coarticulation in speech recognition, optical character recognition for hand-written manuscripts, or word disambiguation in natural language processing. 

RNNs have been the most common architecture for dealing with sequential data, at least until the appearance of transformers. In their gated version, i.e. LSTMs and GRUs, they have been used with some success in the implementation of trading systems \cite{chen2015}\cite{nelson2017}. However, RNN training is problematic and limited when they have to deal with long-term dependencies.

\item
{\bf Convolutional neural networks [CNNs]} are widely-used in signal processing applications, including speech recognition and computer vision (e.g., image classification, object detection and tracking...). They are based on the discrete convolution operation:
$$ y(t) = ( x \star w ) (t) = \sum_i x(i) w(t-i)$$
whose first operator $x$ is the input and whose second operator $w$ is often referred to as the kernel or convolution mask. In neural networks, the kernel or mask $w$ can be learnt using backpropagation.

In CNNs, there are no recurrent connections. A network topology with weight sharing substitutes for the recurrent connections of RNNs. Whereas RNNs receive their input sequentially (one value at a time), CNNs receive their input in parallel (the whole sequence at once).

Time delay neural networks [TDNNs] \cite{lang1988} were proposed to classify patterns with shift-invariance (i.e. they do not require explicit segmentation prior to classification) and model context at each layer of the network. They perform a 1D convolution across time and they were originally proposed for speech recognition \cite{waibel1989}\cite{lang1990}, where the automatic determination of precise segments or feature boundaries was difficult or impossible.

In computer vision applications, 2D convolutions are used. Their historical antecedent is Fukushima's Neocognitron \cite{fukushima1979} \cite{fukushima1980}, inspired, at least partially, by the architecture of the first layers of our visual cortex. 

CNNs have been successfully applied to solve all kind of problems involving signals, no matter whether they are one-dimensional (sounds, time series, sequences...) or multidimensional (images and video). They have also been used for stock market prediction, as in CNNpred \cite{hoseinzade2019}.

Multi-scale CNNs \cite{sermanet2011} are an interesting CNN variation and were originally proposed for traffic sign recognition (i.e. image classification). Conventional CNNs are organized in strict feed-forward layered architectures, in which the output of one layer is fed
only to the following layer. Multi-scale CNNs add skip connections, connecting the output of each convolutional layer to the input of the final fully-connected network layers, which are responsible for the final prediction. The motivation for combining representation from multiple stages in the classifier is to provide different scales of the CNN receptive fields to the final classifier.

\item
{\bf Attention mechanisms and transformers} \cite{vaswani2017} also eliminate the recurrent connections in RNNs, without arbitrarily restricting connections between neurons in the nework as CNNs do.

Transformers convert their sequential input into a numerical representation, a sequence of token vectors. 
An embedding layer converts tokens and positions of the tokens into vector representations.
At each transformer layer, tokens are contextualized within the scope of the context window with other tokens via a parallel multi-head attention mechanism that allows the signal for key tokens to be amplified and the signal for less important tokens to be attenuated.
Transformer layers carry out repeated transformations on the vector representations, extracting more and more information by alternating attention and feed-forward layers.

The components of the Transformer attention mechanism are two vectors of dimension $d_k$, the query $q$ and the key $k$, and a third vector of dimension $d_v$, the values $v$. The matrices $Q$, $K$, and $V$ pack sets of queries, keys, and values. Three projection matrices $W_Q$, $W_K$, and $W_V$ generate different subspace representations of the query, key, and value matrices. Finally, the projection matrix $W_O$ is used for the multi-head attention output. The output of the attention mechanism is computed as a weighted sum of the values, where the weight assigned to each value is computed by a compatibility function of the query with the corresponding key. The scaled dot-product attention computes
$$ attention(Q,K,V) = softmax \left( \frac{Q K^T}{\sqrt{d_k}} \right) V $$
where each output of the $softmax(x)$ function is obtained from
$$ y_j = \frac{e^{x_j}}{\sum_i e^{x_i}}$$
In essence, the attention function is a mapping between a query and a set of key-value pairs to an output. 

Transformers led to the development of large language models [LLMs], such as OpenAI's generative pretrained transformers [GPTs], Google's BERT [Bidirectional Encoder Representations from Transformers], and many other alternative models, both open-source and proprietary (see \url{https://huggingface.co/spaces/lmsys/chatbot-arena-leaderboard}). LLMs support popular applications such as ChatGPT or DALL·E. These LLMs are the largest artificial neural networks ever trained, with hundreds of billions of parameters, and they require huge datasets and huge computational resources for training them.

\end{itemize}

Let us know turn our attention to regularization. The goal of regularization methods is improving the performance of deep learning models on unseen data, i.e., data different from those available in the training set. Since neural networks are universal approximators \cite{cybenko1989} \cite{funahashi1989}\cite{hornik1989}\cite{montufar2014}, they can easily overfit training data and their performance may suffer when deployed and used on novel data.

Training data, apart from the relevant patterns we are interested in, also contains noise. Accidental regularities due to the particular dataset used for training and sampling errors might leak into the models we train. When we are training a model parameters, however, we cannot separate relevant from accidental regularities, so we always incur the risk of overfitting. As John von Neumann said, with four parameters you can fit an elephant, with five you can make him wiggle his trunk. In deep learning models, you can have thousands, millions, and even billions of parameters. 

The combined use of multiple regularization methods is more than advisable to to prevent overfitting when training deep learning models. Many techniques can help us achieve this goal. For instance, we can obtain more training data. This is often the best option if we have the capability to train a neural network using more data and the ability of obtaining those extra data, which is not always the case. We can also try to adjust the network parameters so that the neural network has the right capacity for our particular problem, enough for identifying the regularities in our data that are truly relevant, but not too much, so that it cannot learn spurious patterns (assuming, obviously, that the spurious ones are somehow weaker than the relevant ones). A third viable strategy is creating an ensemble using multiple individual models. An ensemble can be built using ``model averaging'' by mixing many different models with different hyperparameters or even the same kind of model trained on different subsets of training data, a technique known as bagging. An ensemble can also be built using Bayesian fitting, a probabilistic approach based on combining the predictions of multiple neural networks with the same architecture but different weight matrices.

A wide variety of heuristic regularization techniques can be applied during the training process of a neural network. In their essence, they all try to limit the capacity of the neural network, so that it fits the complexity of the problem the network is trying to model. In particular, the following regularization methods can be used and combined when training a deep learning model:

\begin{itemize}

\item
{\bf Early stopping} consists of stopping the training process before the neural network overfits data. Stochastic gradient descent is an iterative optimization algorithm used to train neural networks. We start training a network whose parameters are initialized with small weights. As the network is trained, the optimization algorithm makes the model fit the training data better with each iteration. If we keep a separate validation set, we can periodically check the error of the model on that validation set and stop training as soon as overfitting makes its appearance (i.e., when the error on the training data keeps reducing but the error on the validation set starts to increase).

\item
{\bf Loss function regularization} adds additional terms to the loss function we optimize when training a neural network, so that error minimization is not the only criterion used to adjust the network parameters. There are different forms of this kind of regularization:

\begin{itemize}

\item 
L2 regularization \cite{weigend1990}, also known as weight decay \cite{gupta1998}, Tikhonov regularization \cite{tikhonov1977}, or ridge regression \cite{hoerl1970}, adds a quadratic penalty term to the loss function, the Euclidean or L2 norm:
$$ L = L_{error} + \lambda \frac{1}{2} \sum_k w_k^2$$
where $L_{error}$ is the loss function corresponding to the model error (e.g., MSE for regression, cross-entropy for classification), and $\lambda$ is a regularization factor (another hyperparameter to be adjusted).

Weight decay prevents overfitting and leads to softer models, whose outputs change more slowly when their inputs change. When a network has two similar inputs, e.g. two heavily-correlated signals, L2 regularization will prefer to share weight between them, instead of assigning all the weight to a single input.

\item
L1 regularization \cite{ishikawa1995}\cite{kozma1996}, also known as Lasso regression \cite{tibshirani1996}, resorts to the L1 norm, the norm used for the Manhattan distance:
$$ L = L_{error} + \lambda \sum_k | w_k |$$
Lasso is actually an acronym for `Least Absolute Shrinkage and Selection Operator.' L1 regularization leads to sparser models, so that many network weights are zeroed. In fact, it was originally proposed as a technique for enabling neural networks to forget.

\item
Elastic nets \cite{zou2005} combine both L1 and L2 regularization. L1+L2 regularization leads to the following loss function in elastic nets:
$$ L = L_{error} +  r \lambda \sum_k | w_k | + (1-r) \lambda \frac{1}{2} \sum_k w_k^2$$
where another hyperparameter $r$ splits the regularization factor between L1 and L2 regularization.
\end{itemize}

\item
{\bf Noise} can also have a regularizing effect. Even when the presence of noise in the training data might lead to overfitting, explicitly adding noise to the training process can also lead to more robust models. Noise can be introduced at different points during training: 

\begin{itemize}

\item
We can add noise to the input data \cite{sietsma1991}, a technique also employed by denoising autoencoders \cite{vincent2010}. In some cases, training with noise is equivalent to L2 regularization \cite{bishop1995}.

\item
We can add noise to the network parameters, i.e., its weights. Weight noise, also known as synaptic noise, has been used to train recurrent neural networks \cite{jim1996}. Under some circumstances, when using zero-mean Gaussian noise, it can also be interpreted as adding a regularization term proportional to $ || \nabla_w y||^2 $ so that small perturbations in the network weights cause a limited effect in the network output $y$ \cite{hochreiter1994}.

\item
Noise can also be added to the activation levels of hidden units \cite{poole2014}. This strategy is similar to the introduction of noise in the input, yet now we ensure that the noise reaches all hidden layers in a deep neural networks.

\item 
Finally, we could even add noise to the gradient itself, i.e., the signal used to train the network \cite{neelakantan2015}.
\end{itemize}

Regardless of the particular strategy used, noise leads to smoother models, whose outputs do not change too much when slight perturbations are present in their input or, maybe, their parameters. It is no surprise that some connections can be established between noise regularization and function loss regularization \cite{chandra2003}.

\item
{\bf Dropout} \cite{srivastava2014} is a particularly effective regularization technique. For each training example, each hidden neuron is randomly ignored (or dropped out) with some probability. For instance, if $p=0.5$, half of the units in the hidden layers are not employed for each particular example (a different randomly-chosen half for each different example). The idea behind dropout is that hidden units stop depending or trusting too much on the work performed by other hidden units in the same layer \cite{krizhevsky2012}. When a hidden unit knows that other units might do its job, it might become a free rider. In Geoffrey Hinton's words, complex conspiracies are not robust! When a hidden unit has to work properly with many different sets of units in the same network layer, it is much more probable that the unit learns something useful.

Dropout regularization is interesting because, in some sense, provides a way to combine multiple models. In fact, it is like building a whole ensemble at the cost of training a single network \cite{wardefarley2014}. For each training example, we sample from a family of $2^H$ different network architectures, a family composed of members who share their weights. Given a particular training dataset, you actually only train some of the models in that exponential size family and each trained model receives just a single training example, an extreme form of bagging. By sharing weights, all models are regularized, even better than with L2 or L1 regularization. In fact, weight sharing is  another regularization technique that partially explains the success of convolutional neural networks [CNNs].

It should be noted that Monte Carlo dropout \cite{gal2016}, used to perform multiple predictions from a single trained model, can help us obtain a more reliable prediction and even an estimate of the uncertainty in our prediction.

\item
{\bf Batch normalization} \cite{ioffe2015} normalizes the layers' inputs by re-centering and re-scaling. Initially proposed to mitigate the problem of internal covariate shift, when parameter initialization and changes in the distribution of the inputs of each layer affect the learning rate of the network, it has been observed to smooth the optimization landscape, allowing for faster training and providing an additional regularizing effect \cite{santurkar2018}.

Batch normalization standardizes not only the network inputs, but the inputs for every network layer. Therefore, it can be interpreted as as preprocessing mechanism for the inputs of each network layer. For each mini-batch, we estimate its mean $\mu$ and variance $\sigma^2$. Next, we standardize using z-scores. Given the activation levels $h$ of a hidden unit for a mini-batch of $m$ examples, its batch-normalized version, $h_{bn}$, is computed as

\begin{align*}
\mu & = \frac{1}{m} \sum_{i=1}^m h_i \\
\sigma^2 & = \frac{1}{m} \sum_{i=1}^m ( h_i - \mu )^2 \\
h_{bn} & = \frac{h - \mu}{ \sqrt{\sigma^2 + \epsilon} }
\end{align*}
where $\epsilon$ is a very small value (e.g., $\epsilon=10^{-8}$) just employed for avoiding divisions by zero. Once the network is trained, the values of $\mu$ and $\sigma^2$ can be replaced by measurements obtained during the training process (e.g. the means and deviations observed for the whole training set). 

As described above, batch normalization could affect the network behavior. For instance, if we normalize all the inputs of a sigmoidal unit, we might be forcing it to operate on its linear regime. So an additional linear transformation is appended to the normalization above: $\gamma h_{bn} + \beta$. This transformation, if we used $\gamma=\sigma$ and $\beta=\mu$, would result in the original activation levels. The final result is, therefore, $BN_{\gamma, \beta} (h) = \gamma h_{bn} + \beta$, where $\gamma$ and $\beta$ can be learnt as any other network parameters just by using the proper error gradients. In other words, batch normalization preprocesses input data for every network layer and that preprocessing is elegantly integrated into the network training algorithm.

\end{itemize}

\subsection{A Configuration Nightmare: AutoML to the Rescue}

The ACCI database contains thousands of time series. Even after discarding individual stocks and ETFs, we still have hundreds of time series that can be relevant for predicting trends in financial markets. 

Feature or variable selection, time series selection in our case, is a key issue when building ML models. When we have $v$ variables, there are $2^v$ subsets of them we use as input to a predictive model. With hundreds of variables, we can easily reach the Eddington number, the estimated number of protons in the observable universe, which is similar to the estimated number of atoms, around $10^{80}$. We can even surpass a googol, $10^{100}$. The difference factor, $10^{20}$, might not seem too large, but it is one million times the World GDP in US dollars (above \$100 trillion dollars, i.e. $10^{14}$ dollars). World population is below $10^{10}$ people, which is still four times larger than the average number of seconds in a lifetime.

Some traditional ML techniques require feature or variable selection in order to be effective. More advanced learning algorithms, such as deep learning models, are able to do feature selection automatically (though better results can often be achieved with the proper data preprocessing work). Some regularization methods help us obtain sparser models, hence effectively discarding the less relevant inputs. Even then, there are still multiple degrees of freedom in the design of a ML model.

Apart from variable or feature selection (i.e., choosing individual model inputs), we have a wide range of techniques at our disposal for preprocessing them. From selecting the right normalization or standardization strategy to choosing their proper representation, including a multitude of automated feature extraction and dimensionality reduction techniques.

Model outputs can often be modeled in different ways, without a clear indication of which of them might behave better for our particular prediction problem.

As we discussed previously, ML models also have their own hyperparameters, whose interaction cannot often be predicted. In the case of deep learning models, this problem is exacerbated. We have a wide range of neural network architectures. Layers within the network can be connected in multiple ways. The size of each layer and the particular details of each unit within each layer can also be varied.

Apart from the final model hyperparameters themselves, additional degrees of freedom are available for the model training process. The optimization algorithms behind learning techniques have their own parameters, from learning rates and momentum to weight initialization strategies. And, of course, multiple regularization methods can be mixed and matched at will, each one with its own hyperparameters.

The resulting number of model training configurations, which grows exponentially with the number of hyperparameters, is mind-boggling. How can be choose the most suitable hyperparameters for a particular predictive model?

\begin{itemize}

\item
{\bf Grid search} performs an exhaustive search by testing every possible combination of values for the different hyperparameters. Given $d$ hyperparameters, if the i-th hyperparameter can take $v_i$ different values, grid search performs $\prod_{i=1}^d v_i$ experiments. In the simplest case, for binary hyperparameters, grid search leads to $2^d$ tests, which grows exponentially with the number of dimensions of the search space (i.e., the number of hyperparameters). Given its prohibitive cost, grid search can only be performed for a very limited number of hyperparameter values, typically as a final fine-tuning strategy.

\item 
{\bf Greedy strategies} are much more efficient. For instance, we can test for different values of a particular hyperparameter to check which value leads to better results. Using this specific values, we then test different values for the following hyperparameter, and so on. That strategy requires $\sum_{i=1}^d v_i$ experiments for each cycle through the different algorithm hyperparameters. For binary hyperparameters, it requires $2d$ tests, and effort that is linear with the number of dimensions in the search space. It is basically a hill-climbing strategy and, as such, can be stuck at a local optimum. It is like coordinate descent, but without any kind of optimality guarantee.

\item
{\bf Random search} \cite{bergstra2012} can avoid local optima and is often a more effective strategy. Given a fixed budget, we just perform a set of experiments for random hyperparameter configurations. This strategy is very easy to implement and often leads to better results. It is more effective in practice than the aforementioned systematic search alternatives. Why? Because, for the same number of experiments, it explores a wider zone of the search space.

\end{itemize}


Whereas greedy strategies exploit the best known result around a specific search area, random search explores different areas within the search space. Intelligent search strategies can also be used to achieve a reasonable trade-off between exploration and exploitation. Instead of systematically exploring a particular set of hyperparameter values or sampling those values at random, we can guide the search process. Machine learning techniques can be used for that and they have led to the development of the AutoML field, the automation of ML:

\begin{itemize}
\item 
{\bf Bayesian optimization} \cite{mackay1992} methods maintain a probabilistic belief about the performance of the model in terms of its hyperparameters. They use a so-called acquisition function to determine which experiment to perform next. Bayesian optimization is particularly well-suited to hyperparameter optimization problems, since the function we are trying to optimize is an expensive black-box function. Evaluating a particular combination of hyperparameter values requires running an expensive training process, yet the evaluation of the acquisition function is much cheaper.

Bayesian optimization employs an underlying Gaussian process model.
A Gaussian process models a distribution over functions. When you sample a Gaussian process, you obtain a particular function. That function can be used to estimate the expected improvement for different combination of hyperparameter values, and then guide the search process. For instance, we might start by randomly sampling the search space, with a few samples per dimension (e.g., $3d$). Those samples are used to train a Gaussian process model, which is repeatedly sampled to propose the next hyperparameter configuration to test. The process is repeated iteratively, updating the Gaussian process model each time we get new results on actual models.

Since the Gaussian process, used as acquisition function, is inexpensive with respect to training a whole deep learning model and is commensurate with how desirable a model might be, we optimize the acquisition function to select the configuration for the next experiment. We have just replaced our original optimization problem with another optimization problem, but on a
much-cheaper function.

A common alternative, GP-UCB \cite{srinivas2010} \cite{kandasamy2016}, where UCB stands for upper confidence bound, is used by popular hyperparameter optimizers such as Keras Tuner, Spearmint \cite{snoek2012}, or auto-sklearn \cite{feurer2015} \cite{feurer2019} \cite{feurer2022}.
The GP-UCB acquisition function contains explicit exploitation and exploration terms. 
Under certain conditions, the iterative application of this acquisition function will converge to the true global optimum of the actual function we are trying to optimize.


\item
{\bf Machine learning techniques} can be used for hyperparameter optimization, in what might be called a second-order ML problem. In other words, we use ML algorithms to learn how to set the parameters of another ML algorithm. Many different techniques have been tried in practice, from random forests in SMAC [Sequential Model-based Algorithm Configuration] \cite{hutter2011} and tree-structured Parzen estimators [TPE] in Hyperopt \cite{bergstra2011} \cite{bergstra2013} to neural networks in DNGO [Deep Networks for Global Optimization] \cite{snoek2015}. Of course, those ML algorithms also have their own hyperparameters, which should be set properly to obtain good results.

\item
{\bf Gradients} can be computed, or merely estimated, to perform hyperparameter optimization. Hypergrad \cite{maclaurin2015} extracts gradients for the whole training process. The computation of hypergradients, i.e., the hyperparameter gradients, is conceptually appealing but not too practical. Derivative-free optimization \cite{conn2009}, used in tools such as the RBFOpt optimizer \cite{diaz2017}, choose random points and approximate the true gradients. Even a crude estimate can help us choose in which direction to move within the extremely large space of hyperparameter configurations.

\item
{\bf Freezing heuristics} can help us reduce the use of computational resources in hyperparameter optimization. Freeze-thaw Bayesian optimization \cite{swersky2014} monitors current experiments so that it is able to decide when to launch a new experiment, when to freeze an experiment that no longer seems promising, and when to thaw a previously-frozen experiment when the discovery of additional information makes it promising again. Lazy hyperparameter tuning \cite{zheng2013} aims to determine when a particular hyperparameter configuration will not lead to promising results. In that case, we can avoid its costly evaluation. In some sense, it freezes a particular configuration before it even exists. In principle, we could use any suitable heuristic for that purpose, even use some kind of statistical criterion, such as an hypothesis test, as the following family of hyperparameter optimization techniques.

\item
{\bf Racing algorithms} \cite{birattari2002} are based on statistical testing and lead to early stopping-based hyperparameter optimization methods.
Alternative configurations are repeatedly evaluated throughout the race. Whenever statistical significant evidence about the inferiority of an alternative is found, that alternative will be dropped out of the race. That is, racing algorithms focus the search on the most promising configurations using statistical tests to discard the ones that perform poorly. It can be guaranteed that the eventual survivors will be statistically significantly better than the discarded ones. The irace package \cite{lopez2016} implements the iterated racing algorithm. Successive halving [SHA] \cite{jamieson2016} begins as a random search and periodically prunes low-performing models. Asynchronous successive halving [ASHA] \cite{li2020} improves SHA by removing the need to synchronously evaluate and prune low-performing models. The Hyperband optimizer \cite{li2018} is a higher level early stopping-based algorithm that invokes SHA or ASHA multiple times with varying levels of pruning aggressiveness.

\item
{\bf Neuroevolution} hybridizes artificial neural networks with evolutionary computation, leading to EANNs [Evolving Artificial Neural Networks] \cite{downing2015}. This field provides an endless source of neat acronyms. Neuroevolution techniques employ evolutionary algorithms to create artificial neural networks. Some of them use direct encodings for the parameters in a neural network (e.g. their topology), such as SANE [Symbiotic Adaptive Neuroevolution], ESP [Enforced Sup-Populations], NEAT [NeuroEvolution of Augmenting Topologies], or CGP [Cartesian Genetic Programming]. Other neuroevolution algorithms are based on indirect encodings, when the genes used to represent neural networks in the evolutionary algorithms encode rules that can later be used to build t henetworks, such as CE [Cellular Encoding], G2L [Graph Grammar L-Systems], HyperNEAT [Hypercube-based NEAT] and ES-HyperNEAT [Evolvable-
Substrate HyperNEAT], or MENA [Model of Evolving Neural Aggregates].

Evolutionary optimizers \cite{kousiouris2011} \cite{miikkulainen2024} and population-based training [PBT] \cite{jaderberg2017} \cite{li2019} can be used to learn both hyperparameter values and network weights. For instance, in a genetic algorithm such as NSGA-II \cite{deb2002}, multiple learning processes operate independently, using different hyperparameters. With evolutionary methods, poorly performing models are iteratively replaced with models that adopt modified hyperparameter values and weights based on the better performers. Current neuroevolutionary techniques can be viewed as the descendants of the neural network pruning and growing algorithms that proliferated in the 1990's but were gradually displaced by the adoption of multiple regularization techniques, which were much more efficient from a computational point of view.

\end{itemize}


Hyperparameter tuning and optimization is an active area of research. Apart from the popular approaches based on Bayesian optimization using Gaussian processes or those derived from evolutionary computation (e.g., the genetic algorithms used by EANNs), many other strategies are currently under investigation, from spectral techniques \cite{hazan2017} and Lipschitz functions \cite{malherbe2017} to a myriad of metaheuristics, of which genetic algorithms are just an instance.



Even though many of the proposed approaches start with lofty goals, they frequently settle on relatively humble results. And they often do so at a huge computational cost, since many strategies are not truly scalable. Sometimes, however, you get interesting results. Some AutoML systems have been able to propose neural network architectures for solving specific problems that human experts deemed not to be suitable for that kind of problems. In some sense, AutoML systems are able to discover things we do not know, with surprising results now and then.

Even when taking into account its limitations, AutoML is still much better that performing tests by hand. That is not the kind of tasks we, humans, do well. AutoML techniques avoid undesirable psychological biases, something we cannot do, no matter how hard we try. AutoML is less prone to work well with the methods we like and worse with the ones we dislike. Its unrelenting determination and more objective approach is invaluable for solving complex ML problems.

\subsection{Model Evaluation}

There is still a question we have not answered for the problem we are trying to solve, that of predicting trends in financial markets. From the countably infinite number of potential models and hyperparameter configurations, how do we choose the right one?

The incorrect method would be testing as many alternative as possible to check which one performs better on the test set. That would be easy to do but would led to a misleading estimation of how well our model will perform in practice. The particular model that works better on our particular test set might not be the one what would perform better on {\em other} test sets (or the future data we will apply our model to).


A better approach involves the use of a validation set. We then have to split the available data into three different parts: a training set (to learn the model parameters, e.g. neural network weights), a validation set (not used during training, to be used to decide which hyperparameter configuration is the most suitable), and a test set (to obtain an unbiased estimate of how well our model will work in the real world).


Cross-validation [CV], sometimes called rotation estimation or out-of-sample testing, provides an even better model validation technique for assessing how a particular hyperparameter configuration will generalize to an independent data set. Cross-validation provides an estimate of the quality of a predictive model and also of the stability of its parameters.


Cross-validation includes resampling and sample splitting methods that use different portions of the data to test and train a model on different iterations. In $k$-fold cross-validation [$k$-CV], the original data set is randomly partitioned into $k$ equal-sized subsets, often referred to as folds. 10-fold cross-validation [10-CV] is commonly used. Of the $k$ folds, one is retained as the validation data for testing the model, and the remaining $k-1$ folds are used as training data. The process is repeated $k$ times, with each of the $k$ folds used exactly once as the validation data, to obtain $k$ estimates of the model performance. The $k$ results can then be averaged to produce a single estimation. 


\begin{figure}[!t]
\centering
\includegraphics[width=3in]{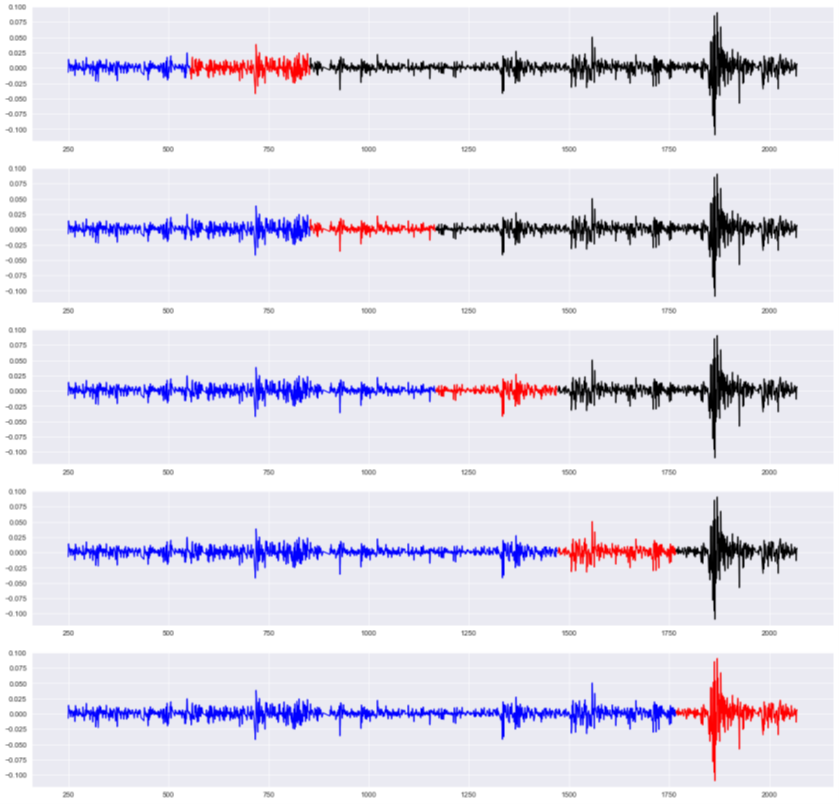}
\caption{Walk-forward cross validation, from \cite{fernandez2022}: Models are trained on historical data (in blue) and tested using test sets (in red) that always posterior to the data used to train them.}
\label{fig-cv}
\end{figure}

In time series data, the independence assumption of the random partition performed by the standard cross-validation procedure is violated. Current time series values are supposed to depend on their past values, i.e., $x_{t+1}$ depends on $x_t$, and so on. Therefore, we should never use values in the training set that are posterior in time to the values on the test set we use to evaluate model performance. That would lead to biased overly-optimistic estimates for our true model performance. 

Walk-forward Validation [WFV] is a time-series cross-validation technique used to assess the performance of predictive models for time-ordered data. This includes trend prediction and time series forecasting for stock prices, weather data, sales figures, and financial markets. When the temporal sequence matters in our prediction, test sets must always be posterior than training sets. 

Walk-forward cross-validation proceeds as follows. Let us assume that we are using natural years to split our time series. You start by training a model on historical data up to the year $y$ and testing it on the year $y+1$. Then, you train another model on data up to the year $y+1$ to test it on the year $y+2$. And so on. At the end, you will have $k$ unbiased estimates of the true performance on your model, which you can use to decide which hyperparameter setting will likely perform better in practice.

Walk-forward cross-validation respects the temporal order of observations, making it suitable for time-series data (temporal consistency). Since the model is never trained on future data, the risk of data leakage is minimized. Obviously, it requires enough data to train models on historical data for different training windows. Its unbiased estimates can then provide a realistic assessment of how models will perform on future, unseen data. 

\subsection{Inside the Black Box: The Limits of XAI Techniques}

Even though they are powerful, deep learning models are often criticised for being black boxes. Their full behavior cannot be understood in the same sense that we can easily grasp the behavior of a linear models or decision trees. Linear models have a limited number of parameters, just one for each input, and their output is derived directly from changes in their inputs (in fact, the output change is proportional to the input change and that proportion is given by the corresponding input weight in the model). Decision trees are symbolic models and their decisions can be easily followed, from the root of the tree to a leaf, just by looking at the criteria used to split the tree. However, deep learning models comprise thousands, millions, or even billions of parameters we cannot really know how they interact. This has led to the recent surge of eXplainable AI [XAI] techniques.

XAI focuses on reasoning behind the decisions or predictions made by ML models so that they can be made more understandable and transparent. Interpretability is another common term used to refer to this goal. In black box ML models, nobody can explain why the model arrived at a particular conclusion. So XAI can be crucial for practical and even regulatory reasons.

Tim Miller \cite{miller2019} has explored some key issues in XAI. First, explanations are contrastive. They are sought in response to particular counterfactual cases, a.k.a., foils. People do not ask why P happened, but rather why P happened instead of Q. Second, explanations are selected in a biased manner. People rarely, if ever, expect an explanation that consists of an actual and complete cause of an event. Humans are adept at selecting one or two causes and this selection is influenced by cognitive biases. Third, probabilities probably do not matter. Referring to probabilities or statistical relationships in explanation is not as effective as referring to causes. The most likely explanation is not always the best explanation for a person. Using statistical generalizations to explain why events occur is unsatisfying, unless accompanied by an underlying causal explanation for the generalisation itself. Last but not least, explanations are social. They are presented relative to the explainer’s beliefs about the explainee’s beliefs. Explanations are not just the presentation of associations and causes (causal attribution), they are contextual. While an event may have many causes, often the explainee cares only about a small subset (relevant to the context), the explainer selects a subset of this subset (based on several different criteria), and explainer and explainee may interact and argue about this explanation.

When working with ML models, model interpretability can refer to the degree to which an observer can understand the cause of a decision \cite{biran2017} or he degree to which a human observer can consistently predict the model output. \cite{kim2016}.


Whereas some XAI techniques have been proposed for particular kinds of ML models (e.g., saliency maps highlight which regions of an image lead to its final classification), the most interesting ones are model-agnostic. Model-agnostic methods can be used with any ML technique, so they are general-purpose XAI tools:

\begin{itemize}

\item
A {\bf partial dependence plot} [PDP], or PD plot, shows the marginal effect one or two features have on the predicted outcome of a machine learning model. 
The partial dependence function computes averages in the training data to provide, for each value of the chosen feature, what the average marginal effect on the prediction is. It has a causal interpretation: by changing a feature and measuring the changes in the predictions, we analyze the apparent causal relationship between the feature and the prediction \cite{zhao2021}.

\item
The {\bf individual conditional expectation} [ICE] is the local method equivalent to the global PDP. Global XAI methods describe the average behavior of a ML models, while local XAI methods explain individual predictions. ICE plots display how each particular prediction changes when a feature changes \cite{goldstein2015}. Variants of this method include centered ICE [c-ICE] plots and derivative ICE [d-ICE] plots.

\item
{\bf Accumulated local effects} [ALE] plots show how features influence the prediction of a ML model on average \cite{apley2020}. While M plots average the predictions over the conditional distribution, ALE plots average the changes in the predictions and accumulate them.

\item
{\bf Permutation feature importance} measures the importance of a feature by calculating the increase in the model's prediction error after
shuffling its values in the training set. A feature is deemed to be important if shuffling its values increases the model error; i.e., the
model relied on the feature for performing the right predictions. A model-agnostic version of feature importance is called model class reliance \cite{fisher2019}. 

\item
{\bf Feature interaction} techniques describe how different features interact with each other in situations when predictions cannot be expressed as the combination of individual feature effects and reductionism can no longer work. Feature interaction can be measured using Friedman's H-statistic \cite{friedman2008}, variable interaction networks [VIN] \cite{hooker2004}, or partial dependence-based feature interaction \cite{greenwell2018}.

\item
{\bf Anchors} \cite{ribeiro2018}, a.k.a., scoped rules, explain the individual predictions of black box models by finding decision rules that anchor the model prediction. A rule anchors a local prediction when changes in features not considered by the rule do not affect the final prediction.

\item
{\bf LIME} stands for {\bf Local Interpretable Model-agnostic Explanations} \cite{ribeiro2016}. Local surrogate models are trained to approximate the predictions of the underlying black box model. For instance, we can train a linear model in the neighborhood of a particular example so that linear model can be used as an local approximation to the behavior of the more complex, nonlinear model. The surrogate model should provide a good local approximation to the the black box model, but it does not have to be a good global approximation. This property is called local fidelity. LIME just  trains local surrogate models to explain individual predictions.

\item
{\bf SHAP} stands for {\bf Shapley Additive exPlanations} \cite{lundberg2017}. Yet another XAI method to explain individual predictions, SHAP is based on Shapley values \cite{shapley1953}, proposed by the Nobel Prize winner Lloyd S. Shapley in his studies of coallitional game theory. Shapley values assign unique payouts to players depending on their contribution to the total payout generated by the collaboration of all players. Shapley values result in a linear model where the contribution of the $j$-th feature to the prediction is computed as the average marginal contribution of a feature value across all possible coalitions. Shapley values can be estimated using Monte Carlo sampling \cite{strumbelj2014}, since only approximate solutions are feasible. The interpretation of the estimated Shapley value is fairly reasonable: the contribution of a feature value to the difference between the actual prediction and the mean prediction given the current set of feature values.
SHAP has a solid theoretical foundation and its original proposal, KernelSHAP, connects LIME with Shapley values. However, it is slow, ignores feature dependencies, and it can be misinterpreted. 

\end{itemize}

Due to their current popularity, you can easily find surveys \cite{burkart2021} and even textbooks \cite{molnar2021} on XAI techniques.

Some commercial vendors offer their own implementation of XAI techniques such as LIME or SHAP, often at a premium. It should be noted that the explanations they provide are unstable \cite{alvarez2018} and that they can also hide biases \cite{slack2020}. In fact, explanations can be manipulated on purpose. Both LIME and SHAP, for instance, can be used to create intentionally misleading interpretations.

Given a particular example, the explanation provided by a local XAI technique might hold. However, for a different example that also fits that explanation, a black box ML model might provide a completely different conclusion. Of course, the XAI method will eagerly provide a novel explanation to describe why the conclusion is now different, even the opposite. In some sense, therefore, local XAI methods give excuses more than explanations.


\section{Conclusion}

This whitepaper describes the approach we use to predict trends in financial markets with the help of deep learning techniques. 

If the efficient-market hypothesis [EMH] held, asset prices would always reflect all the available information and consistently beating the market would be an impossibility. 

If we just use historical prices for the asset price we want to predict, as most traditional forecasting techniques do, our predictive accuracy would be modest. In the limit, for efficient markets, our accuracy would be 50\%. Even without using additional information, effective algorithmic trading strategies have been developed that can provide an edge.

By taking additional context into account, we can improve our odds of beating the market. Our models use of time series instead of the scalar values more traditional techniques employ as input (e.g., hand-crafted features or trading alphas). This poses significant challenges from the technical point of view, given the large number of degrees of freedom you have when designing deep learning models, from feature selection, engineering, and extraction, to hyperparameter optimization and fine tuning.

The incorporation of hundreds of time series, not even hundreds of scalar variables, is an unsurmountable limitation of many traditional ML techniques, so they cannot benefit from the deluge of available data at our fingertips. Traditional solutions based on risk indicators also tend to rely on linear models. Since we live in a complex nonlinear world, linear approximations are not always applicable. Linear models are not only less accurate than deep learning models, but they are also less robust to the presence of noise in data and, more importantly, they are less responsive when market trends change, which they often do abruptly.

Deep learning models provide quantitative advantages with respect to other ML techniques for dealing with the complexity of current financial markets, hence they are at the core of ACCI risk indicators.

\bibliographystyle{ieeetr}
\bibliography{bibliography}

\begin{IEEEbiography}[{\includegraphics[width=1in,height=1.25in,clip,keepaspectratio]{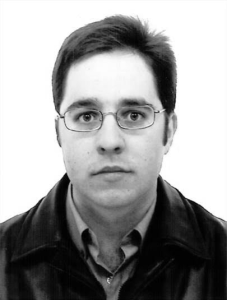}}]{Fernando Berzal}
received his PhD degree in Computer Science from the University of Granada, Spain, in 2002 and he was awarded the Computer Science Studies National First Prize by the Spanish Ministry of Education, in 2000. He is an associate professor with the Department of Computer Science and Artificial Intelligence, University of Granada, Spain, and he has been a visiting research scientist at the University of Illinois at Urbana-Champaign. His research interests include model-driven software development, software design, and the application of data mining techniques, as well as Artificial Intelligence, complex networks, and deep learning. He is a senior member of the ACM and also a member of the IEEE Computer Society.
\end{IEEEbiography}

\begin{IEEEbiography}[{\includegraphics[width=1in,height=1.25in,clip,keepaspectratio]{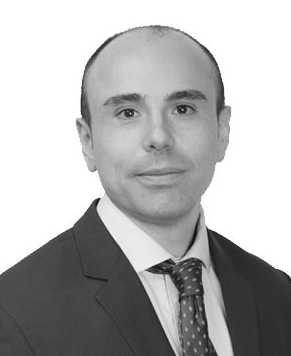}}]{Alberto Garc\'\i a}
holds a Business \& Administration Degree and has also obtained some of the most recognized Investment Certificates in Finance. During his career, Alberto has earned the CFA, CAIA, FRM, ESG CFA Investment Certificate, and the IMC designation. He worked from 2005 to 2010 as a Senior Business Analyst for Ahorro Corporación and moved to UK, where he worked as an Investment Analyst/Quantitative Developer for Collidr until 2015. From 2015 to November 2022, he worked for Santander Asset Management first in the Portfolio Construction and Risk Management Team to move internally to the Asset Allocation team where he spent the last 3 years. He is currently the head of global asset allocation at ACCI Capital Investments.

\end{IEEEbiography}

\end{document}